\documentclass[reprint,aps,prl, floatfix,showpacs]{revtex4-1}

\usepackage[utf8]{inputenc}
\usepackage{graphicx}
\usepackage{bm}
\usepackage{color}

\begin{document}

\title{Structure of the cholesteric-isotropic interface}
\date{\today}
\author{N. R. Bernardino}\email{nelsonrb@cii.fc.ul.pt}
\author{M. C. F. Pereira}
\author{N. M. Silvestre}
\author{M. M. \surname{Telo da Gama}}
\affiliation{Centro de F\'{\i}sica Te\'{o}rica e Computacional, Faculdade de Ci\^{e}ncias, Universidade de Lisboa, 1749-016 Lisboa, Portugal.}
\affiliation{Departamento de F\'{\i}sica, Faculdade de Ci\^{e}ncias, Universidade de Lisboa, 1749-016
Lisboa, Portugal.}

\begin{abstract}

The interface of a cholesteric liquid crystal with an isotropic fluid can display a range of unusual properties, such as a layer of topological defects close to an undulated interface. These properties have been know for a long time and have been explored for technological applications as a tunable substrate for colloidal self-assembly. However, from a fundamental point of view, this interface remains poorly understood and even basic properties, such as the dependence of the surface tension on the attributes of the liquid crystal, remain unknown. Here, we present a systematic calculation of the structure and surface tension of the cholesteric-isotropic interface and how these vary with the properties of the liquid crystal. We also suggest the intriguing possibility of wetting of this interface by a blue phase.   

\end{abstract}


\maketitle

Liquid crystals (LCs) illustrate many of the most fascinating concepts of fundamental Statistical Physics, having an intermediate order between the isotropic liquids and the highly symmetric solids \cite{deGennes1993,Wright1989}. Thanks to their central role in the display industry nematic phases, the simplest and the least ordered LCs, are widely studied from the fundamental, and also from the applied, point of view.

\begin{figure}[htp]
 \centering
 \begin{center}
\includegraphics[width=\columnwidth]{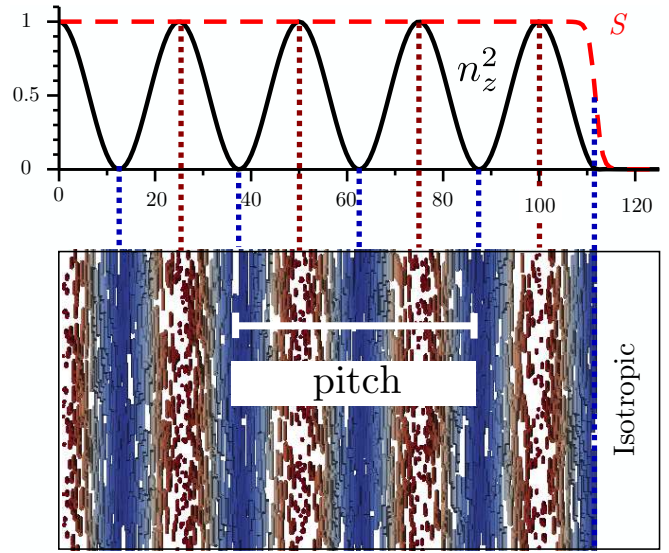}
\end{center}
 \caption{\label{fig_basic}(color online) Cholesteric-isotropic interface with planar anchoring. The cholesteric has a pitch $P=50\xi$ and the local orientation of the LC is color coded from in-plane (blue) to out-of-plane (red). Also plotted are the values of the scalar order parameter $S$ (red dashed line) and the out-of-plane orientation of the LC $n_z^2$ (black line) along the horizontal direction. The interface is identified by the abrupt change in the value of $S$ from $S_b\approx1$ to $0$.}
\end{figure}

Nematic LCs have no positional but posses orientational ordering, i.e., a preferred direction along which the molecules align. In cholesterics, also known as chiral nematics, this preferred direction rotates in space due to the chirality of the constituents, or of a dopant, describing a helix with a given pitch (see Fig.~\ref{fig_basic}). This extra length scale and extra degree of ordering in one dimension are responsible for the peculiar properties of the cholesteric phases, which exhibit many of the features of layered systems.

Such properties also manifest themselves at the interfaces of cholesterics which display a range of unusual features, such as non-planar interfaces and nucleation of a layer of topological defects  \cite{Cladis1972,Saupe1973,Meister1996a,Meister1996,Zola2013,Lintuvuori2013}. These phenomena have been described experimentally and theoretically \cite{Cladis1972,Saupe1973,Meister1996a,Meister1996}, as well as their implications to wetting phenomena \cite{Zola2013}. It is however remarkable that even basic interfacial properties, such as the surface tension, and their dependence on the important parameters of the system, the pitch and the elastic constants, are not know. This surprising fact is due to the complex structure of the interface that is not amenable to analytical results; even the numerical analysis is non-trivial as it requires adequate descriptions on very different length and time scales. 

In this Letter we describe the properties of the interface of a cholesteric LC with the isotropic liquid phase, using a Landau-de Gennes (mean field) free energy description. This method is particularly suitable to study the cholesteric-isotropic interface as it describes simultaneously both phases, the transition between them, and the intrinsic structure of the interface. We show how fundamental interfacial properties such as the structure of the interface and the surface tension depend on the parameters of the system. The calculation of the surface tension requires a careful discussion of what is the interface and how this definition depends on the length scale at which we are observing it. We also reveal the intriguing possibility of wetting of the cholesteric-isotropic phase by a more ordered blue phase \cite{Wright1989}. Our focus is on the fundamental interfacial physics of the cholesteric-isotropic interface, however the ordered layer of topological lines that nucleate at the interface 
has potentially interesting technological applications, as a tunable substrate for colloidal templating \cite{Lintuvuori2013}.

The starting point of our analysis is the Landau-de Gennes free energy for a cholesteric \cite{deGennes1993, patricio2011}, which is built from the usual expansion in the lowest order terms of the symmetric, traceless, tensor order parameter $Q_{ij}$, split into elastic $f_{e}$, and bulk terms $f_{b}$, $F = \int d V \left( f_{e} + f_{b}\right)$. Here $f_{e} = \frac{1}{3+2\kappa} ( Q_{ij,k} Q_{ji,k} +  4q_0 Q_{il}\epsilon_{ijk} Q_{kl,j} + 4q_0^2 Q_{ij}Q_{ji} +\kappa Q_{ij,j}Q_{ki,k})$ and $f_{b} = \frac{2}{3}\tau Q_{ij}Q_{ji} - \frac{8}{3}Q_{ij}Q_{jk}Q_{ki} + \frac{4}{9}(Q_{ij}Q_{ji})^2$, where summation over repeated indices is assumed. We use a dimensionless free energy. $\kappa = L_2/L_1$ is the ratio of two elastic constants $L_1$ and $L_2$; $\kappa=0$ is equivalent to the usual one-constant approximation in the Frank-Oseen formalism \cite{deGennes1993}. $q_0$ is the inverse wavelength of the cholesteric pitch $P = \frac{2\pi}{q_0}$. In 
this model the nematic phase is described by the limit of infinite pitch or, equivalently, the limit $q_0\to 0$. $\tau$ is a reduced temperature whose value determines the equilibrium bulk phase. While for a nematic $\tau=1$ is the coexistence temperature between the nematic and the isotropic phases, for a cholesteric the coexistence temperature depends both on $\kappa$ and the pitch.
All lengths are measured in units of the correlation length $\xi$, which is the scale of the typical size of the topological defects and of the width of the LC-isotropic interface. As a reference, for the nematic LC 5CB the correlation length at room temperature is around $15nm$ so $P=1000\xi$ is equivalent to $P=15\mu m$. The order in the LC phase is described by a scalar order parameter $S$ with values $S=0$ in the isotropic phase and $S=S_b$ in the LC phase. For a nematic, at the coexistence temperature, $S_b=1$.  For cholesterics, however, $S_b$ depends weakly on $\kappa$ and $P$.
We assume translational invariance along the $z$ direction and thus calculate the configuration on the $xy$ plane. The Landau-de Gennes free energy is minimized using the Finite Element Method with a relaxation scheme, through the commercial program COMSOL 3.5a (http://www.comsol.com). The meshes used are such that the precision of our numerical  results is better than $1\%$. To ensure a good resolution of the interface we use a finer mesh of maximum size $\xi$ close to it.

The first step to understand the cholesteric-isotropic interface is to look at the configuration of the interface when the interface favors planar anchoring, i.e. the local direction of the LC is parallel to the plane of the interface, as in Fig.~\ref{fig_basic}. In this case the layered structure of the bulk cholesteric is not disturbed by the interface. The value of the scalar order parameter $S$ changes at the interface but, since this change occurs over a distance comparable to the bulk correlation length $\xi$, we expect that if the pitch $P\gg\xi$ the cholesteric-isotropic interface is similar to the nematic-isotropic interface, with a similar value of the surface tension. In fact, a calculation of the surface tension using a simple ansatz for the interface gives a value identical to that of the nematic-isotropic interface \cite{deGennes1971,Patricio2008}:
$\sigma_\| = \frac{1}{6}\sqrt{\frac{6+\kappa}{3+2\kappa}}$.
We tested numerically the accuracy of the ansatz and the result for $\sigma_\|$, and found very good agreement, within the accuracy of the numerical results. Recall that for the nematic similar calculations show that the equilibrium orientation at the interface can only take two values, depending on $\kappa$. If $\kappa>0$ the anchoring is planar, as just described. On the other hand, $\kappa<0$ favors homeotropic (perpendicular) anchoring, with the LC oriented parallel to the normal to the interface and a surface tension $\sigma_\perp=\frac{\sqrt{2}}{6}$. Other values of the orientation of the LC at the interface are not possible, in equilibrium, within this Landau-de Gennes free energy.

\begin{figure*}[htp]
 \centering
 \begin{center}
\includegraphics[width=\textwidth]{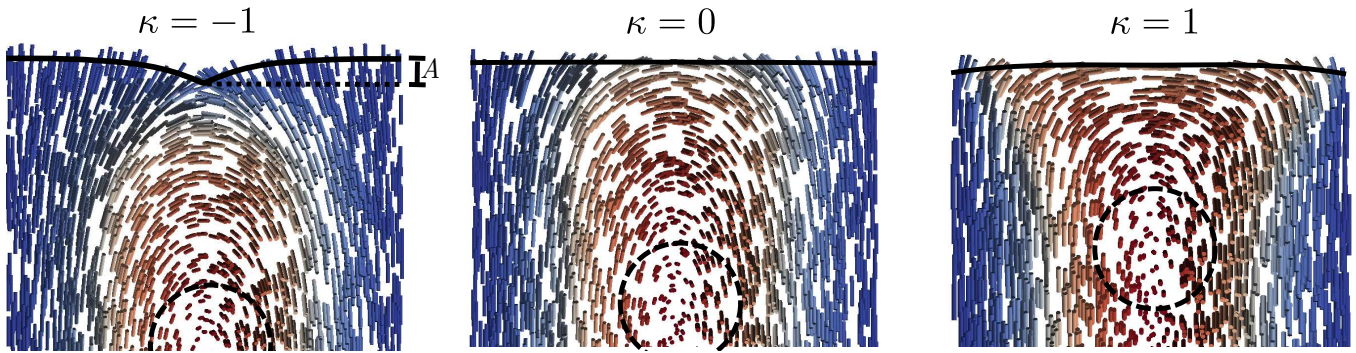}
\end{center}
 \caption{\label{fig_configs} (color online) Configurations of the interface for $P=1000\xi$, (left) $\kappa = -1$, (middle) $\kappa=0$, and (right) $\kappa=1$. Notice that the orientation of the LC at the interface goes from mostly homeotropic ($\kappa = -1$) to mostly planar ($\kappa=1$) as the sign of $\kappa$ changes. The orientation of the cylinders depicts the orientation of the liquid crystal and the color is proportional to the out-of-plane orientation of the LC (red for out-of-plane; blue for in-plane). The black line marks the position of the interface at $S=S_b/2$ and the dashed circle identifies the $\lambda^+$ disclination line. The amplitude of the undulations of the interface $A$ is defined as the difference between the heights of the maximum and minimum of the line $S=S_b/2$. The lateral size of each configuration is $P/2=500\xi$. The configuration for $\kappa=1$ (right) is metastable.}
\end{figure*}

For a cholesteric with $\kappa<0$, favoring interfacial homeotropic anchoring, the differences from the nematic are striking as there is intrinsic frustration in the system: it is not possible to find a configuration that, at the same time, has homeotropic anchoring at the interface and is consistent with the layered structure of the cholesteric. As a consequence, the equilibrium configuration includes both the creation of topological defects ($\lambda^+$ disclination lines \cite{deGennes1993}) and undulations of the interface \cite{Cladis1972,Saupe1973,Meister1996a,Meister1996,Zola2013}. This can be seen in Fig.~\ref{fig_configs} that illustrates the configuration of the interface with a pitch $P=1000\xi$ and $\kappa = -1, 0, 1$: the interface, defined by the line with $S = S_b/2$, is undulated and there is a non-singular topological defect close to it. Taking $\kappa=0$ as a reference, where neither planar nor homeotropic anchoring is preferred, it is clear that for $\kappa=-1$ the anchoring is mostly 
homeotropic and 
for $\kappa=1$ the anchoring is mostly planar, as expected from the results for the nematic-isotropic interface.

To obtain these solutions we started with an initial configuration of a bulk cholesteric that changes abruptly to an isotropic phase perpendicular to the ``layers''. The system is then allowed to evolve towards the minimum of the free energy. Note that the configurations for $\kappa \ge 0$ are metastable or even unstable: the equilibrium configuration has the ``layers'' parallel to the interface. We also note that for high values of $\kappa$, which strongly favor planar anchoring at the interface, the system can satisfy the anchoring requirement, while still twisting in the direction parallel to the interface, by breaking the symmetry near the interface and doubling the periodicity of the configuration from $P/2$ to $P$. Even though this configuration of the interface is metastable it is probably observable in experiments and is related to some configurations that have been described before \cite{Saupe1973, Meister1996a}.

Given the complex distortions close to the interface, the definition of what is the interface and the corresponding surface tension is somewhat subtle. There can be two, apparently opposed, points of view on what constitutes the interface and how to calculate the surface tension. Perhaps the most intuitive one is: for $P\gg\xi$ the cholesteric-isotropic interface is similar to the nematic-isotropic interface, including the value of the surface tension. 
We used this reasoning to justify why the surface tension of a cholesteric with planar alignment at the interface, as in Fig.~\ref{fig_basic}, has the same value as the nematic-isotropic surface tension. From this point of view the undulation of the interface and the distortions of the cholesteric layers are not part of the interface and should not be included in the calculation of the surface tension. This reasoning is adopted implicitly in calculations such as the ones in Ref.~\cite{Meister1996a}.

On the other hand, at a macroscopic scale the interface is effectively flat. Notice that the amplitude of the undulation in Fig.~\ref{fig_configs} is a small fraction of the pitch. For pitches on the micron scale the undulations of the interface are most likely below the visible wavelengths and so the interface will appear flat when observed with an optical microscope. From a thermodynamic, macroscopic, point of view the surface tension is the free energy cost of increasing the area of an interface. This means that if we double the area of the interface, we must double the number of distortions. Hence, the energetic contributions of the undulations and of the distortions of the cholesteric layers must be included in the free energy. They are an integral part of the interface.

\begin{figure}[htp]
 \centering
 \begin{center}
 \includegraphics[width=\columnwidth]{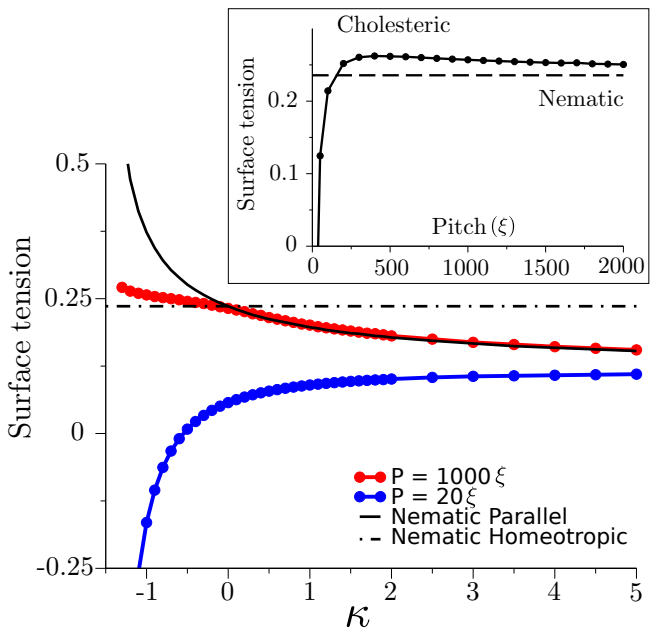}
\end{center}
 \caption{\label{fig_surftension} (color online) Surface tension with $\kappa$ and pitch of a cholesteric-isotropic interface with the layers perpendicular to the interface. (main, bottom) Surface tension with $\kappa$ for pitch $P=1000\xi$ (red) and $P=20\xi$ (blue). Also shown  are the results for the nematic with planar (black, full) and homeotropic anchoring at the interface (black, dashed-dotted). The surface tension for $P=20\xi$ increases with $\kappa$, by contrast to what happens for large values of the pitch, and becomes negative for $\kappa \lesssim -0.5$ signaling that the cholesteric phase is no longer the thermodynamically stable phase in this region of parameters. If the system is allowed to relax further a blue phase nucleates from the interface, Fig~\ref{fig_blue_phase}. (inset, top) Surface tension with pitch for a cholesteric with $\kappa=-1$ (full line) and for a nematic with homeotropic anchoring at the interface (dashed horizontal line). The surface tension reaches a value close to the 
asymptotic for $P>200\xi$, with a maximum at around $P=400\xi$ and then decreases slowly to the value of the surface tension of the nematic-isotropic interface with homeotropic anchoring.}
\end{figure}

Hence the definition of the interface and the surface tension depends on the length scales at which we look at the system. We are interested in the thermodynamic surface tension and so adopt the latter view. To be a bit more quantitative we can calculate the surface tension of the configurations for given pitch and $\kappa$. Since the free energy is constructed such that at coexistence the free energy of the LC and that of the isotropic phase are equal and correspond to the ground-state, the thermodynamic surface tension is simply the volume integral of the free energy of the configuration that minimizes the free energy, such as those shown on Fig.~\ref{fig_configs}. The calculation of the surface tension of a cholesteric with $P=1000\xi$ reveals a surprisingly small effect of the undulations of the interface and of the cholesteric layers on the surface tension, as seen in Fig.~\ref{fig_surftension}. For $\kappa>0$ the orientation of the layers perpendicular to the interface does not minimize the free 
energy and this configuration is metastable or even unstable. The  equilibrium configuration has the layers arranged parallel to the interface as in Fig.~\ref{fig_basic}. For $\kappa<0$ there are some deviations from the surface tension of the nematic but, again, of a surprisingly small magnitude given the increased surface area of the interface (at small length scales) and of the distortions of the cholesteric layers. 

\begin{figure}[htp]
 \centering
 \begin{center}
 \includegraphics[width=0.9\columnwidth]{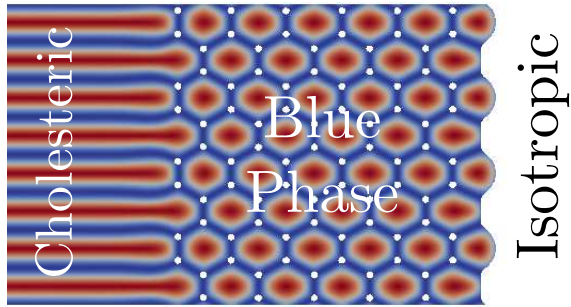}
\end{center}
 \caption{\label{fig_blue_phase}Blue phase at the cholesteric-isotropic interface. For small values of the pitch a blue phase condenses from the interface. This raises the intriguing possibility of wetting of the cholesteric-isotropic interface by a blue phase.}
\end{figure}

Also shown are the results for a much smaller value of the pitch, $P=20\xi$. In this case the behavior is considerably different: the surface tension is clearly different from that of a nematic, decreasing as we move to more negative values of $\kappa$ and it becomes negative  at $\kappa \approx -0.5$. The latter indicates that in this region of the parameters the cholesteric phase is no longer the thermodynamically stable phase and is a signature of the presence of stable blue phases in the phase diagram \cite{Wright1989}. This strong decrease of the surface tension for negative values of $\kappa$ is due to the formation of regions of double-twist close to the interface (precursors of the blue phase), that have lower free energy than the bulk cholesteric, Fig.~\ref{fig_blue_phase}.

In the inset of Fig.~\ref{fig_surftension} we plot the surface tension for several values of the pitch and $\kappa=-1$. We find a strong dependence of the surface tension on the value of the pitch for $P<200\xi$, reflecting the effect of the double twist regions and of the proximity to the blue phases in the bulk phase diagram. Only for $P>200\xi$ does the result approach the asymptotic limit of the nematic-isotropic interface.

\begin{figure}[htp]
 \centering
 \begin{center}
 \includegraphics[width=\columnwidth]{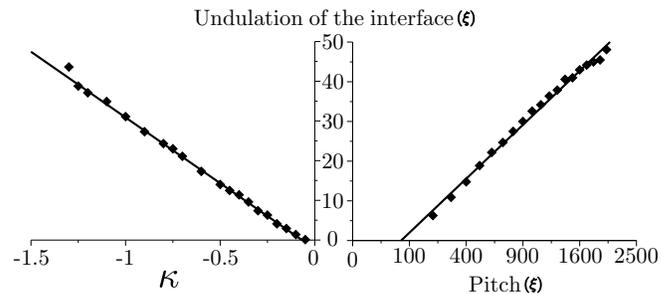}
\end{center}
 \caption{\label{fig_amplitude}Amplitude of the undulations of the interface with $\kappa$ for $P=1000\xi$ (left) and with the square root of the pitch for $\kappa=-1$ (right). The straight lines are fits to the results: $A=-33.1\kappa-2.2$ (left) and $A=1.359\sqrt{P}-11.6$ (right).}
\end{figure}

Let us now look at the dependence of the amplitude of the interfacial undulation $A$, measured as the difference between the heights of the maximum and minimum of the isoline $S=S_b/2$ (Fig.~\ref{fig_configs}), on both the pitch and $\kappa$, plotted in Fig.~\ref{fig_amplitude}. The interfacial undulations increase linearly with $|\kappa|$, for $\kappa<0$. For $\kappa>0$ the equilibrium state is the configuration with the layers parallel to the interface and thus the undulation of the interface is zero. The same qualitative behavior was found for other values of the pitch, $P>200\xi$. Fig.~\ref{fig_amplitude} also reveals the scaling $A\sim\sqrt{P}$. In the limit of infinite pitch (a nematic) the ratio $A/P$ goes to zero, as expected. We observed the same qualitative behavior for other values of $\kappa<0$.

The explanation of the dependence of the amplitude of the undulations with $\kappa$ and the pitch is not trivial. The undulations are the result of a delicate interplay between the elastic distortions of the liquid crystal, the surface tension and the anchoring at the interface. Calculations of the free energy of an interface that may change its configuration at fixed director configuration, using the ansatz of Ref.~\cite{Meister1996a}, results in the predictions $A\sim P$ and $A\sim \kappa^2$. Comparing our results with the configurations that result from this ansatz it is clear that the failure, of the ansatz, may be traced to the inaccurate description of the  configuration of the LC close to the interface. The fact that the scaling $A \sim P$ breaks down is a clear sign of the presence of two length scales: $P$ and $\xi$. Despite that in general $\xi\ll P$ the influence of $\xi$ is important as the interface undulates to avoid the nucleation 
of a second topological defect, whose size and position, are determined by $\xi$. The scaling of $A$ with $\kappa$ and $P$ is determined by the minimization of the distortions while avoiding the nucleation of this topological defect, which we have not been able to obtain analytically.

Finally, we come back to the results for smaller values of the pitch. We have already seen the signs of the presence of blue phases in the phase diagram: negative surface tension of the cholesteric-isotropic interface. In these systems we observed nucleation of the blue phase at the interface, Fig.~\ref{fig_blue_phase}. The condensation of a phase from an interface proceeds without nucleation barriers and thus the growth is more ordered and with fewer defects than those that result from condensation in the bulk. This provides an interesting route to assemble defect free blue phases, either in experiments or in simulations. In the latter case the formation of a blue phase usually requires the tunning of the initial conditions of the liquid crystal, something that might not always be easy. It is also interesting to note that if we set the parameters of the system at the triple point, at coexistence of the isotropic, cholesteric and blue phase, our preliminary results indicate wetting of the isotropic-
cholesteric interface by the blue 
phase. Since blue phases have the symmetry of solid lattices but flow like liquids the wetting properties of this triple point are bound to be exotic. We plan to explore this in the near future.

To conclude, we showed how the detailed structure and the surface tension of the cholesteric-isotropic interface vary with the important parameters of the liquid crystal. Our approach, based on the Landau-de Gennes free energy, is ideally suited to study the intrinsic structure of the interface and the surface tension.  Past experience suggests that Landau-de Gennes theory captures most of the qualitative phenomena of liquid crystals and is very often surprisingly quantitatively correct, despite the usual shortcomings of mean-field theories when dealing with fluctuations \cite{deGennes1993}. As such, we believe that the main results of our Letter are qualitatively correct and a careful examination of the properties of the cholesteric-isotropic interface should lead to determination of the dependence with the pitch and the validation of our results. This is also the first step in the study of wetting properties of cholesteric liquid crystals. There has been some recent interest in this phenomenon \cite{
Zola2013} but the behavior at patterned surfaces is completely unknown, despite the fundamental interest of wetting of patterned substrates by liquid crystals \cite{Patricio2008,patricio2011}.

Our focus was on the fundamental properties of the cholesteric-isotropic interface but our results might have interesting technological applications. Recently it was shown that it might be difficult to use the array of disclination lines at a cholesteric-oil interface for colloidal templating, due to the huge number of metastable states that can form \cite{Lintuvuori2013}. Since the undulations of the interface are much bigger for the cholesteric-isotropic interface, and the surface tension much lower, the energy landscape is very different and colloidal templating might be easier at such interfaces.

\acknowledgments
We acknowledge financial support from the Portuguese Foundation for Science and Technology (FCT) under Contracts nos. PTDC/FIS/119162/2010, PEst-OE/FIS/UI0618/2014, SFRH/BPD/63183/2009, and 
EXCEL/FIS-NAN/0083/2012. We also acknowledge helpful discussions with R. Zola and L. Evangelista.

%

\end{document}